\documentclass{ptephy_v1}%%%%%% to generate preprint number with ptep logo

\preprintnumber{XXXX-XXXX} %%% %%% Insert preprint number here

\begin{document}

\title{Structure of double pionic atoms}

%%%% To generate auto affiliation numbers please use \author{}\affil{} command

\author{Akari Tani}
\affil[1]{Department of Agricultural, Life and Environmental Sciences, Tottori University, Tottori 680-8551, Japan}
%\affil{Insert first author address here \email{xxxx@xxxx.ac.jp}}

\author[1,2,*]{Natsumi Ikeno}
\affil[2]{Departamento de F\'{\i}sica Te\'orica and IFIC, Centro Mixto Universidad de
Valencia-CSIC, Institutos de Investigaci\'on de Paterna, Aptdo.22085,
46071 Valencia, Spain }

\author[3]{Daisuke Jido}
\affil[3]{Department of Physics, Tokyo Institute of Technology, Meguro, Tokyo 152-8551, Japan}

\author[4,5]{Hideko Nagahiro}
\affil[4]{Department of Physics, Nara Women's University, Nara 630-8506, Japan}
\affil[5]{Research Center for Nuclear Physics (RCNP), Osaka University, Ibaraki 567-0047, Japan}%

\author[3]{Hiroyuki Fujioka}

\author[6]{Kenta Itahashi}
\affil[6]{Nishina Center for Accelerator-Based Science, RIKEN, 2-1 Hirosawa,
Wako, Saitama 351-0198, Japan
\email{ikeno@tottori-u.ac.jp}}

\author[4]{Satoru Hirenzaki}

% \author[4]{Insert last author name here\thanks{These authors contributed equally to this work}}
% \affil{Insert last author address here}

%%% To include the collaborator name... Please use the command "\collaborator"
%%% For example: \collaborator{ATLAS Collaboration}

\begin{abstract}%
We study theoretically the structure of double pionic atoms, in which two negatively charged pions ($\pi^-$) are bound in the atomic orbits. 
The double pionic atom is considered to be an interesting system from the point of view of the multi bosonic systems.
In addition, it could be possible to deduce valuable information on the isospin $I = 2$ $\pi\pi$ interaction and the pion-nucleus strong interaction. 
In this paper, we take into account the $\pi\pi$ strong and electromagnetic interactions, and evaluate the effects on the binding energies by perturbation theory for the double pionic atoms in heavy nuclei. We investigate several combinations of two pionic states and find that the order of magnitude of the energy shifts due to the $\pi\pi$ interaction is around 10~keV for the strong interaction and around 100~keV for the electromagnetic interaction for the ground states.

\end{abstract}

\subjectindex{xxxx, xxx}

\maketitle

\section{Introduction \label{intro}}
%---- Interest and motivation
The structure and formation of the meson-nucleus bound systems, such as mesic atoms and mesic nuclei, have been studied for a long time since they are considered to be one of the most interesting objects to investigate the meson-nucleus interactions and the in-medium meson properties~\cite{Batty:1997zp,Friedman:2007zza}.  
%----- pionic atom
In particular, the information on the partial restoration of chiral symmetry at finite nuclear density was successfully obtained by determining a pion-nucleus optical potential parameter using the precisely measured pionic 1$s$ state binding energies in Sn isotopes~\cite{Piatom,KSuzuki}.  Thus, the deeply bound pionic atom is thought to be one of the best systems to deduce the precise information on meson-nucleus interaction in the nuclear medium.  Theoretical discussions based on the symmetry of the strong interaction support these studies of the chiral symmetry in medium using the observables of pion-nucleus systems~\cite{Kolo,Jido,Friedman:2019zhc}.

The experimental searches for various meson-nucleus bound systems have been performed, recently. 
The spectroscopic study of the pionic atoms in the $^{122}$Sn($d, ^3$He) reaction was carried out at RI beam facility in RIKEN~\cite{Nishi}.
Another experiment is already planned to measure the deeply bound pionic atoms by $^{112, 124}$Sn($d, ^3$He) reactions to deduce the isotope dependence of the in-medium pion properties using Sn isotopes~\cite{proposal2019}.  
In addition, the search for the  $\eta'$ mesic nucleus by the $^{12}$C($p,d$) reaction was performed at GSI to investigate the origin of the large mass of  $\eta'$ which is considered to be related to the $U_A$(1) anomaly and the broken chiral symmetry~\cite{Tanaka:2016bcp,Jido:2011pq,Nagahiro:2012aq}.  The $\eta$ mesic nucleus was also studied experimentally using light ions at COSY to investigate the baryon resonance $N^*(1535)$ which is a candidate of the chiral partner of nucleon~\cite{Adlarson:2013xg,Skurzok:2011aa,Adlarson:2016dme,Wilkin:2007aa,Xie:2016zhs,Ikeno:2017xyb,Skurzok:2018paa,Xie:2018aeg}.

In this paper we study double pionic atoms, in which two negatively charged pions ($\pi^-$) are bound to the atomic orbits in one nucleus.
The double pionic atoms have never been measured experimentally so far. 
However, some theoretical studies for the formation of the double pionic atom have been reported around 30 years ago for the ($\pi^-, \pi^+$) and ($\pi^- ,p$) reactions~\cite{formation2pion,Nieves:1992kc}.
They discussed challenging ideas to produce the double pionic atoms and calculated the formation cross sections, even before the deeply bound `single' pionic atoms were discovered experimentally.
It should be very interesting to revisit the study of the double pionic atoms based on 
the latest theoretical and experimental knowledge accumulated through the studies of meson-nucleus 
bound systems as mentioned above.

The double pionic atoms are considered to have interesting and important features as follows.
First, the structure of the double pionic atoms must be affected by the strong interaction between two pions 
in addition to the $\pi\pi$ electromagnetic interaction and the pion-nucleus interaction.  
Thus, we expect to obtain the information on the isospin $I=2$ $\pi\pi$ 
interaction  and even more we could extract the in-medium modification of the 
$\pi\pi$ 
interaction from the study of the spectrum of the double pionic atoms, which will be a new clue to deepen our understandings of the chiral dynamics 
of hadrons.  

In addition, the double pionic atoms considered in this paper could also be a first step on the studies of multi-bosonic atoms, though it could be highly-academic 
to consider the structure of the multi-bosonic atoms.
Since bosonic systems have no exclusion principle, we believe that it will be extremely interesting and exciting to have some scientific insights on the periodic table and more generally the chemistry of bosonic atoms.

This paper is organized as follows. In Sec.~\ref{formula}, we explain the formulation for the study of the structure of single and double pionic atoms 
based on the theoretical methods reported in Refs.~\cite{Piatom,Ikeno,Ikeno:2011aa,Ikeno:2013wza,Ikeno:2015ioa}. 
For the double pionic atoms, we evaluate the energy shifts  of the bound states of two pions by the $\pi\pi$ interaction using the perturbation theory with the realistic wave functions of the single pionic atoms.
In Sec.~\ref{sec:result}, we present the numerical results of the structure of the double pionic atoms.
A summary is given in Sec.~\ref{sec:summary}.

\section{Structure of single and double pionic atoms \label{formula}}
In this section, we explain our theoretical formulation to calculate the energy and 
  the wave function of pionic atoms. The effects of the $\pi\pi$ interaction is treated in the first order perturbation theory and the energy shift induced by the $\pi\pi$ interaction is evaluated by the expectation value of the $\pi\pi$ interaction potential with the unperturbed wave function of the double pionic atom that is a product of the wave functions of the single pionic atom.

\subsection{Structure of single pionic atom \label{signle_pi}}

The structure of the single pionic atoms is studied by solving the Klein-Gordon equation~\cite{Piatom,Ikeno},
\begin{equation}
\left[-\nabla^{2}+\mu^{2}+2\mu V_{\rm{opt}}(r)\right]\phi(\vec{r})
=\left[ E- V_{\rm{FC}}(r)\right]^{2} \phi(\vec{r}),
\label{KGeq}
\end{equation}
where $\mu$ is the pion-nucleus reduced mass, 
$E$ the eigen energy written as $E=\mu-B_{\pi}-\displaystyle \frac{i}{2}\Gamma$ 
with the binding energy $B_\pi$ and the width $\Gamma$ of the atomic states. 
$V_{\rm FC}$ is the Coulomb potential with a finite nuclear charge
density distribution $\rho_{ch}(r)$:
\begin{equation}
V_{\rm{FC}}(r)
= - \frac{e^2}{4 \pi \epsilon_0} \int\frac{\rho_{ch}(r^{\prime})}{|\vec{r}-\vec{r^{\prime}}|} d\vec{r^{\prime}}.
\label{coulom-pot}
\end{equation}
The charge density distribution is written by the Woods-Saxon form as,
\begin{equation}
\rho_{ch}(r) = \frac{ \rho_{ch 0}}{1+ \exp[(r-R_{ch})/a_{ch}]}.
\label{rho_ch}
\end{equation}
The parameters of the charge distributions are taken from
Refs.~\cite{Fricke,238U} and summarized in Table~\ref{R,a} for the nuclei
considered in this paper.
The distribution function $\rho_{ch}(r)$ 
is normalized to the number of the protons in the nucleus. 

For the pion-nucleus optical potential $V_{\rm opt}$ in Eq.~(\ref{KGeq}),
we use the Ericson-Ericson type potential~\cite{Ericson} as,
\begin{eqnarray}
2\mu V_{\rm opt}(r)
= -4\pi[b(r)+\varepsilon_2B_0\rho^2(r)] 
+4\pi\nabla\cdot[c(r)+\varepsilon_2^{-1}C_0\rho^2(r)]L(r)\nabla,
\label{Vopt}
\end{eqnarray}
with
\begin{eqnarray}
b(r) &=& \varepsilon_{1}[b_{0}\rho(r)+b_{1}[\rho_n(r)-\rho_p(r)]], \label{Vopt_Swave1}\\
c(r) &=& \varepsilon_{1}^{-1}[c_{0}\rho(r)+c_{1} [\rho_n(r)-\rho_p(r)]],\label{Vopt_Pwave2}\\
L(r) &=& \left\{ 1+ \frac{4}{3} \pi\lambda[c(r)+
      \varepsilon_{2}^{-1}C_{0}\rho^{2} (r)] \right\}^{-1}, \label{Vopt_Pwave3}
\end{eqnarray}
where $\varepsilon_{1}$ and $\varepsilon_{2}$ are defined as
$\varepsilon_{1}=1+\displaystyle \frac{\mu}{M}$ and 
$\varepsilon_{2}=1+\displaystyle \frac{\mu}{2M}$ with 
the nucleon mass $M$.  
As a standard parameter set, we use the potential parameters in Ref.~\cite{SM}, which are compiled in Table~\ref{table:Vopt_para}. 
 As the possible energy dependence of the pion-nucleus interaction is already renormalized into the potential parameters, the coefficients of the linear density term differ from the corresponding in-vacuum values~\cite{Kolo,Jido}.
The densities $\rho_p$ and $\rho_n$ in Eqs.~(\ref{Vopt_Swave1}) and~(\ref{Vopt_Pwave2}) indicate the distributions of the center of the proton and neutron, which can be deduced from the charge distribution by the prescription described in Ref.~\cite{oset} and can be expressed in the Woods-Saxon form as,
\begin{equation}
\rho_{p, n}(r) =\frac{ \rho_{p,n0}}{1+ \exp[(r-R)/a]}, \hspace{0.5cm}
\rho(r)= \rho_{p}(r) + \rho_{n}(r), 
\label{rho}
\end{equation}
where the parameters $R$ and $a$ are summarized in Table~\ref{R,a}. 
We use the same parameters $R$ and $a$ in the densities of the proton and neutron.
The effects of the possible nuclear deformation of heavy nuclei are expected to be small~\cite{WFnorm1} and are not taken into account in this article.  

\begin{table}[tb]
\caption{\label{R,a}
Radius parameters $R_{ch}$ and diffuseness parameters $a_{ch}$ of the charge distributions in Eq.~(\ref{rho_ch}).
The parameters of $^{122}$Sn and $^{208}$Pb are taken from Ref.~\cite{Fricke}, 
and those of $^{238}$U are from Ref~\cite{238U}.
The diffuseness parameters in Ref.~\cite{Fricke} are fixed to be $a_{ch}=t$/4ln3 for all nuclei with $t=2.30$ fm. The parameters $R$ and $a$ in Eq.~(\ref{rho}) are also shown which are obtained by the prescription in Ref.~\cite{oset}.
}
\centering
\begin{tabular}{cc||c|c|cc} 
\hline
 &~~~nuclide~~~ &~~~ $^{122}$Sn~~~ & ~~~$^{208}$Pb~~~ &~~~ $^{238}$U ~~~ & \\ \hline
 &$R_{ch}$ [fm] & 5.476 & 6.647  & 6.805 & \\ %\hline
 &$a_{ch}$ [fm] & 0.523 & 0.523  & 0.605 & \\ \hline
 &$R $ [fm] & 5.516 & 6.680  & 6.838 & \\ 
 &$a $ [fm] & 0.455 & 0.454  & 0.546 & \\
\hline
\end{tabular}
\end{table}
\begin{table}[tb]
\caption{\label{table:Vopt_para}
Pion-nucleus optical potential parameters~\cite{SM} used in the
 present calculations.}
\centering
\begin{tabular}{cllc}
\hline
 & $b_{0} = -0.0283 m_{\pi}^{-1}$ & $b_{1} = -0.12 m_{\pi}^{-1}$ & \\	
 & $c_{0} =  0.223  m_{\pi}^{-3}$ & $c_{1} =  0.25 m_{\pi}^{-3}$ &  \\	
 & $B_{0} =  0.042i m_{\pi}^{-4}$ & $C_{0} =  0.10i m_{\pi}^{-6}$ &  \\  
 & $\lambda = 1.0$ \\ 
\hline
\end{tabular}
\end{table}

\subsection{$\pi\pi$ interaction}\label{interaction}
To consider the structure of the double pionic atoms,
we take into account the electromagnetic $V^{\rm em}_{\pi \pi}$
and the strong $V^{\rm s}_{\pi \pi}$ interactions between atomic pions.
%--------------------
% Coulom potential
%--------------------
We consider the repulsive Coulomb potential 
with point charge distributions between negatively charged pions 
as the electromagnetic interaction, which is expressed as,
\begin{equation}
 V^{\rm em}_{\pi \pi}(\vec r) = \frac{e^2}{4 \pi \epsilon_0 |\vec r|},
\end{equation}
where $\vec r$ indicates the relative coordinate between two pions.

For the strong interaction $V^{\rm s}_{\pi \pi}$ between pions,
the range of the $V^{\rm s}_{\pi \pi}$ is expected to 
be significantly shorter than that of the pionic atom 
wave function.
Hence, we take a contact interaction for the strong interaction given by the delta function by
\begin{equation}
V^{\rm s}_{\pi \pi}(\vec{r})=V_0 \delta(\vec{r}).
\label{V_delta}
\end{equation}
In order to see the finite range effect, we also take the Gaussian form as,
\begin{equation}
V^{\rm s}_{\pi \pi}(\vec{r})= \frac{V_0}{(b^2 \pi)^{3/2}}
\exp \left[- \frac{\vec{r}^2}{b^2} \right], 
\label{V_gauss}
\end{equation}
with the range parameter $b$.
The value of the $b$ parameter may be expected to be around 0.5~fm in analogy with the $\bar{K} N$ system where $b=0.47$ and $0.66$ fm in Refs.~\cite{KKN1,KKN2}.

To determine the potential strength $V_0$, we consider the $\pi\pi$ scattering length of the isospin-2 ($I=2$) channel obtained in the leading order of the chiral perturbation theory as~\cite{chiral},
\begin{equation}
a_2=-\frac{m_\pi}{16 \pi f_\pi^2}.
\label{a2}
\end{equation}
The value of $a_2$ is evaluated to be $a_2 =-0.045 m_\pi^{-1}$ with the pion decay constant $f_\pi = 93$~MeV.
This value is very close to that determined from $\pi \pi$ data~\cite{pipi_Exp,NA48,GarciaMartin:2011cn}.
The strength of the strong interaction between two $\pi^{-}$ can be connected to
the scattering length $a_2$ of the $I=2$ channel
by the Born approximation as,
\begin{eqnarray}
a_2 &=& \lim_{q \to 0}  
\left\{
-\frac{m_\pi}{4\pi} \int d\vec{r}
e^{-i \vec{q} \cdot \vec{r}} V^{\rm s}_{\pi \pi}(\vec{r})   \right\},
% \nonumber\\
% &=&
% - m_\pi \int r^2 dr V^{\rm s}_{\pi \pi}(\vec{r}),
\label{Born}
\end{eqnarray}
where $\vec{q}$ indicates the momentum transfer between the two pions.
The scattering length $a_2$ is defined to be negative for 
the repulsive interaction $V^{\rm s}_{\pi \pi} > 0$.
From Eq.~(\ref{Born}),
the potential strength $V_0$ for the normalized spatial distributions
in Eqs.~(\ref{V_delta}) and~(\ref{V_gauss}) can be expressed as,
\begin{equation}
 V_0 = - \frac{4 \pi a_2}{m_\pi}= \frac{1}{4 f_\pi^2}.
\label{V0_free}
\end{equation}

We also mention here the effects of the partial restoration of the chiral symmetry and the possible modification of the $V^{\rm s}_{\pi \pi}$ potential.  The effects of the chiral symmetry restoration in-medium, namely the reduction of the order parameter $\langle \bar q q \rangle$ condensate of the chiral symmetry gives rise to the change of the pion decay constant as the function of the nuclear density $\rho$ as~\cite{Jido,KSuzuki}, 
\begin{eqnarray}
 f_\pi^2(\rho) = \left( 1 - 0.36 \frac{\rho}{\rho_0} \right)f_\pi^2,
\label{f_pi_rho}
\end{eqnarray}
with the vacuum value  $f_\pi$ and the normal nuclear density $\rho_0 =0.17$ fm$^{-3}$.
We can simply take into account and investigate this effect in the $\pi \pi$ interaction in the double pionic atoms by replacing the potential strength parameter $V_0$ as, 
\begin{equation}
 V_0(\rho) =  \frac{1}{4 f_\pi^2(\rho)}.
\label{V0_rho}
\end{equation}
Thus, the information on the chiral symmetry restoration
and the reduction of the $f_\pi$ value at finite density
could also be related to the energy shift of the double pionic atoms by the $\pi \pi$ interaction.

\subsection{Structure of double pionic atoms}
In this work, we treat the $\pi \pi$ interaction as perturbation. The unperturbed wave function is given by a product of the wave functions of the single pionic atom which are obtained 
by Eq.~(\ref{KGeq}). 
Since the two pions are identical bosons, the total wave function must be symmetric under the interchange of space and isospin coordinates.
The unperturbed wave function for a double pionic atom is give as,
\begin{eqnarray}
\Phi(\vec{r}_1, \vec{r}_2) = 
%\frac{1}{\sqrt{2}}
N[
\phi_{n \ell m}(\vec{r}_1) \phi_{n' \ell' m'}(\vec{r}_2)  %\nonumber\\
+
\phi_{n' \ell' m'}(\vec{r}_1) \phi_{n \ell m}(\vec{r}_2) 
],
\label{WF_boson}
\end{eqnarray}
where $N$ is the appropriate normalization factor and
$\phi_{n \ell m}$ 
the realistic wave function of the single pionic atom described in Sect.~\ref{signle_pi}.
The one-pion wave function $\phi_{n \ell m}$ is written as  
\begin{equation}
 \phi_{n \ell m}(\vec{r}) = R_{n \ell}(r) \ Y^{m}_{\ell}(\hat{r}),
\end{equation}
with the radial wave function $R_{n \ell}(r)$ and the spherical harmonics $Y^{m}_{\ell}(\hat{r})$.

The unperturbed binding energy and the width of the double pionic states are given as,
\begin{equation}
B_{\pi} = B_{\pi}^{n \ell} +B_{\pi}^{n'\ell'}, 
\label{eq:BE2pi}
\end{equation}
and
\begin{equation}
\Gamma = \Gamma^{n \ell} + \Gamma^{n'\ell'},
\label{eq:gam2pi}
\end{equation}
where we have made use of the fact that
$B_{\pi}, \Gamma \ll \mu$ for the atomic pion 
and adopted the non-relativistic expressions for the eigen energy $E$
obtained by the Klein-Gordon equation.
We mention here that
the width of the double pionic states is
the sum of the single pionic widths
and, thus is around twice larger in average
than that of the single pionic states.

%%%%%%%%%%%%%%%%%%%%%%%%%%%%
Now, we consider the effects of the $\pi \pi$ interaction to the structure of the double pionic atoms. The effects can be evaluated by the perturbation theory for the deeply bound pionic states in heavy nucleus because of the existence of the significantly stronger pion-nucleus interaction.
Since the wave functions of the deeper bound states are expected to have larger overlap, we evaluated the $\pi\pi$ interaction effects for the six different combinations of two pionic states ($n \ell, n'\ell'$), which are $(1s, 1s)$, $(1s,2p)$, $(1s,2s)$, $(2p,2s)$, $(2p, 2p)$ and $(2s,2s)$ states. The pionic $1s$, $2s$ and $2p$ states in Sn were observed experimentally in the spectroscopy of the deeply bound single pionic atoms~\cite{Nishi,Nishi2}.
The wave function of the ground state ($1s,1s$) of the double pionic atom is written as,
\begin{equation}
\Phi_{00} (\vec{r}_1, \vec{r}_2) = \phi_{1 0 0}(\vec{r}_1) \phi_{1 0 0}(\vec{r}_2) .
\label{WF_ground}
\end{equation}
The subscripts of the two pion wave function in L.H.S of Eq.~(\ref{WF_ground}) indicate the orbital angular momentum quantum numbers $(L,M)$ for  two pion system. The wave function of the $(2s,2s)$ state can be expressed in the similar way as Eq.~(\ref{WF_ground}).  
The wave function of the $(1s,2s)$ two pionic state, which also has the total angular momentum $(L,M)=(0,0)$, can be written as, 
\begin{equation}
\Phi_{0 0} (\vec{r}_1, \vec{r}_2) = \frac{1}{\sqrt{2}}[
\phi_{100}(\vec{r}_1) \phi_{200}(\vec{r}_2)  
+
\phi_{200}(\vec{r}_1) \phi_{100}(\vec{r}_2) ].
\label{WF_1s2s}
\end{equation}

For the states with a $s$ state pion and a $p$ state pion, the wave function is written as, 
\begin{equation}
\Phi_{1m} (\vec{r}_1, \vec{r}_2) = \frac{1}{\sqrt{2}}[
\phi_{n00}(\vec{r}_1) \phi_{n'1 m}(\vec{r}_2)  
+
\phi_{n'1 m}(\vec{r}_1) \phi_{n00}(\vec{r}_2) ],
\label{WF_sp}
\end{equation}
where the total angular momentum is $(L, M)= (1,m)$.
The total angular momentum $L$ for the $(2p,2p)$ double pionic states can be $L=0$, 1, and 2, and the wave functions of these states can be expressed using the Clebsch-Gordon coefficient as follows, 
\begin{equation}
\Phi_{L M}(\vec{r}_1, \vec{r}_2) = 
N \{ [ \phi_{21m}(\vec{r}_1) \otimes \phi_{21m'}(\vec{r}_2)]^{L}_{M}  
+ [\phi_{21m'}(\vec{r}_1) \otimes \phi_{21m}(\vec{r}_2) ]^{L}_{M} \},
\label{eq:WF_2p2p}
\end{equation}
where
\begin{equation}
[ \phi_{21m}(\vec{r}_1) \otimes \phi_{21m'}(\vec{r}_2)]^{L}_{M}  
= \sum_{m m'} (1 m 1 m' |L M) \ \phi_{21m}(\vec{r}_1) \phi_{21m'}(\vec{r}_2) .
\label{eq:WF_2p2p_2}
\end{equation}

%%%%%%%%%%%%%%%%%%%%%%%%%%%%%%%%%%%%%%%%%%%%%%%%%
%* energy shift  \\
%%%%%%%%%%%%%%%%%%%%%%%%%%%%%%%%%%%%%%%%%%%%%%%%%
Using these wave functions, the energy shift $\Delta E_{\rm em}$ due to the $\pi\pi$ electromagnetic interaction $V^{\rm em}_{\pi \pi}$ is estimated as,
\begin{eqnarray}
\Delta E_{\rm em} 
= \left\langle \Phi (\vec{r}_1, \vec{r}_2) 
\left| \frac{e^2}{4 \pi \epsilon_0  |\vec{r}_1 - \vec{r}_2|} \right|  
\Phi (\vec{r}_1, \vec{r}_2) \right\rangle,
\label{Eshift_em}
\end{eqnarray}
in the first order perturbation theory.
The energy shift $\Delta E_{\rm s}$ due to the $\pi\pi$ strong interaction $V^{\rm s}_{\pi \pi}$ is estimated in a similar way as,
\begin{eqnarray}
\Delta E_{\rm s}
= \langle \Phi (\vec{r}_1, \vec{r}_2) 
| V^{\rm s}_{\pi \pi} (\vec{r}_1 - \vec{r}_2) |  
\Phi (\vec{r}_1, \vec{r}_2) \rangle, \
\label{eq:Eshift_s}
\end{eqnarray}
where 
$V_{\pi \pi}^{\rm s}$ is defined in Eqs.~(\ref{V_delta})
and~(\ref{V_gauss}).
Explicit expressions of these energy shifts are given in Appendix~\ref{appA}.
There is also a theoretical work of the mesonic-atom-like object for the lighter systems~\cite{Moriya:2019ans}.

Finally, we add a few comments on the orthonormality of the pion wave function.
The pion wave function $\phi(\vec{r})$ in Eq.~(\ref{KGeq}) 
does not have the standard orthonormality
because of the energy dependent Coulomb term $2 E V_{\rm{FC}}(r)$ 
and the imaginary part of the optical potential Im$V_{\rm opt}(r)$,
which make the Hamiltonian energy dependent and non-Hermitian~\cite{WFnorm1,WFnorm2}.
The correction for the non-standard orthonormality
can be evaluated by the correction factors 
$B_{\pi}^{n \ell}/\mu$ and $\Gamma^{n \ell}/\mu$,
and the relative strength of the Coulomb potential
to reduced mass $V_{\rm FC}(r)/\mu$~\cite{WFnorm1,WFnorm2}.
We find that the correction is less than 10\% for Sn region and
can be neglected in this article in the present exploratory level.

\section{Numerical Results} \label{sec:result}
\begin{figure*}[tb]
\includegraphics [width=15cm, height=9cm]{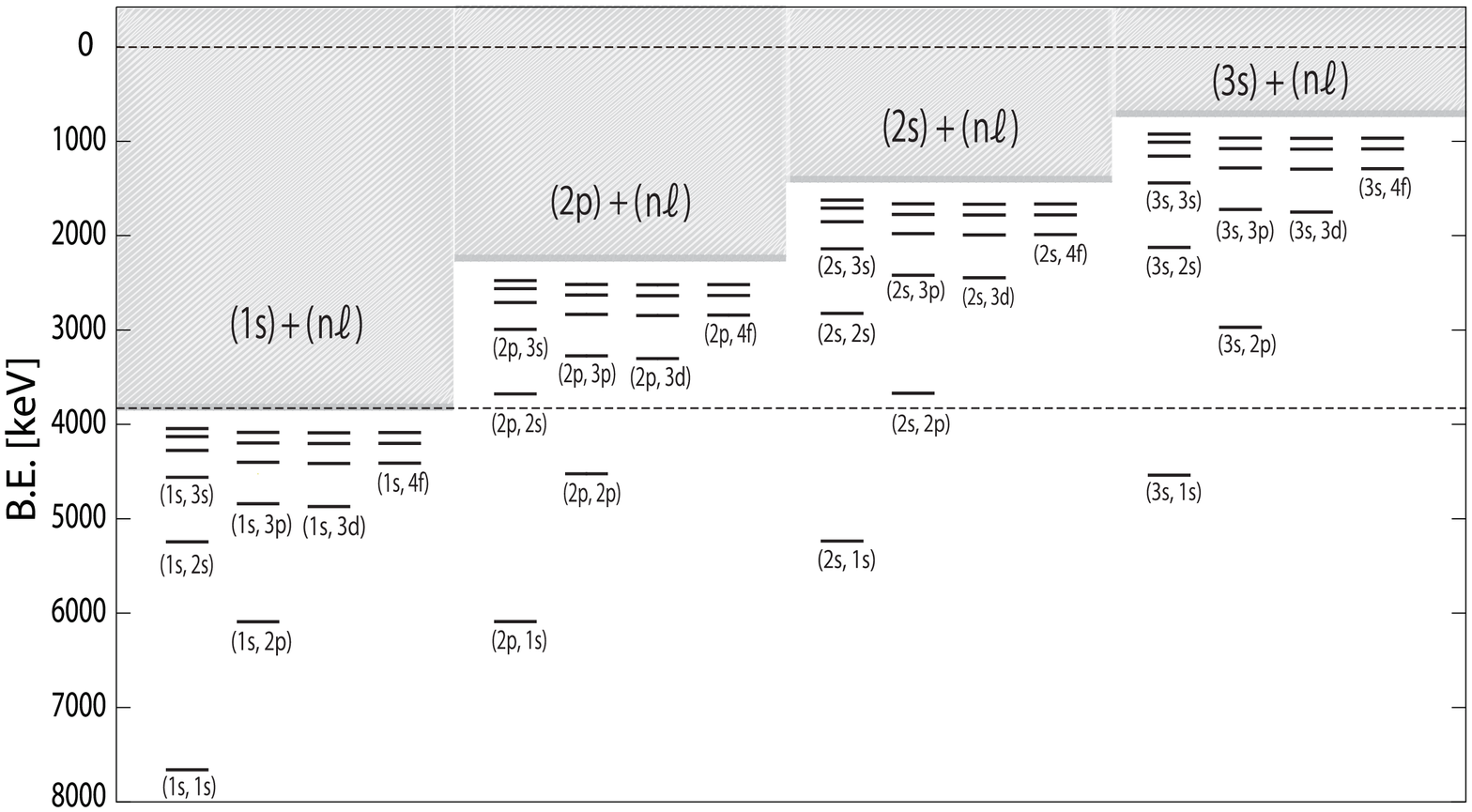}
\caption{Unperturbed energy levels of the double pionic atoms of $^{121}$Sn.
The interactions between pions are not taken into account.
The two dashed lines indicate the threshold energies of the one-pion
and the two-pion quasi-free production, respectively.
The hatched area indicates the energy region of pion continuum levels (see text for details).
\label{level}}
\end{figure*}

In Fig.~\ref{level}, we show the unperturbed energy spectra of the double pionic atoms, which are obtained in Eq.~(\ref{eq:BE2pi}) without $\pi\pi$ interaction. 
In this figure, the energy levels of the double pionic states are shown, where 
the quantum numbers of the state are shown in the form of $(n_1 \ell_1, n_2 \ell_2)$ for two pions.  
Once we fix the state of one pion $(n_1 \ell_1)$ to be the certain state, we have the same energy spectra as the single pionic atoms for possible $(n_2 \ell_2)$ states 
with shifted energies corresponding to the energy of the $(n_1 \ell_1)$ state.  
In Fig.~\ref{level}, we show the levels for four cases with $(n_1 \ell_1) =(1s), (2p), (2s),$ and $(3s) $ states.

%-------------------------------
%Figure: Energy level
%-------------------------------
%-------------------------------

As we naturally expected, the energy spectra of the double pionic atoms are more complicated 
than those of the single pionic atoms.  Especially, we like to mention the existence of two threshold energies in the energy spectrum of two pion state,  which are the quasi-free one-pion production energy and the quasi-free two-pion production energy.
The quasi-free one-pion production threshold energy is defined as $B_{\pi}=B_{\pi}^{1s}+0$ where $B_{\pi}^{1s}= 3850.0$~keV for $^{121}$Sn case~\cite{Ikeno:2015ioa}.
The quasi-free two-pion production threshold energy is $B_{\pi}=0$.
We plot these two threshold energies by dashed lines in this figure.
We can see that there are many discrete states between two thresholds energies.
In the energy region between two thresholds, we will observe the discrete resonance states embedded in the continuum pion spectrum.
By looking at the ($2s, 2p$) states 
as an example, the energy of this state is located above the threshold for the quasi-free one-pion production.
% 1$s$ pion bound state and the other one in the continuum.  
Thus the $(2s, 2p)$ state can decay into $(1s)+$continuum two pionic states in addition to the decay (absorption) modes of the individual pionic states.  
This extra decay mode is known as the autoionization of the excited state of the multi-electronic atoms and emits the Auge electron in the decay process.  
Hence, to investigate the structure of the double pionic atoms precisely, we also need to take into account the coupling of the  discrete excited states and the continuum states.  
%In comparison with ……. than that of helium atoms.  
%----- He
In comparison with the structure of the normal helium atom, we find that we miss a series of spin triplet helium excited states called `orthohelium' in the double pionic atoms because of the symmetric spatial wave functions.
In this sense, the energy spectrum of the double pionic atoms is relatively simpler than that of helium atoms.

\begin{table*}
\centering
\caption{\label{Eshift_table} 
Compilation of the energy shifts of the ground state ($1s, 1s$) of the double pionic atom due to the $\pi\pi$ electromagnetic interaction ($\Delta E_{\rm em}$) and due to the strong interaction ($\Delta E_{\rm s}$) obtained by the first-order perturbation theory.
The results with the $\pi\pi$ interaction with the density dependent
$f_{\pi}$ (Eqs.~(\ref{f_pi_rho}), (\ref{V0_rho})) are indicated as $f_{\pi}(\rho)$.
The results obtained by the Coulomb wave functions ($\phi_{\rm PC}$ and $\phi_{\rm FC}$) are also shown for $^{121}$Sn.
}
%\begin{ruledtabular}
\begin{tabular}{cccccccc}
\hline
\hline
 &   &  & \multicolumn{5}{c}{$\Delta E_{\rm s}$ [keV]} \\ \cline{5-8}
 &  &  & & \multicolumn{2}{c}{$\delta$-function}
 & \multicolumn{2}{c}{Gauss }\\\cline{5-6}\cline{7-8}
Nucleus & Wave function &$\Delta E_{\rm em}$ [keV] &  & $f_{\pi}$ &$f_{\pi}(\rho)$ &  & $f_{\pi}$ \\
\hline
$^{121}$Sn & $\phi_{\rm Opt + FC}$ &  91.0 &  & 7.5 & 7.6 &   & 7.4  \\
           & $\phi_{\rm FC}$ & 154.3 &  & 38.7  & 49.9 & & 38.5   \\
& $\phi_{\rm PC}$ & 232.3 &  & 152.5  & 222.8 & & 150.1    \\         
\hline
$^{207}$Pb & $\phi_{\rm Opt + FC}$ & 97.8 & & 9.4 & 9.7 &  & 9.4   \\
\hline
$^{237}$U  & $\phi_{\rm Opt + FC}$ & 98.4 & & 9.6 &  9.9  & & 9.6  \\
\hline
\hline
\end{tabular}
%\end{ruledtabular}
\end{table*}

\begin{figure}[tb]
\centering
\includegraphics[width=0.6\textwidth]{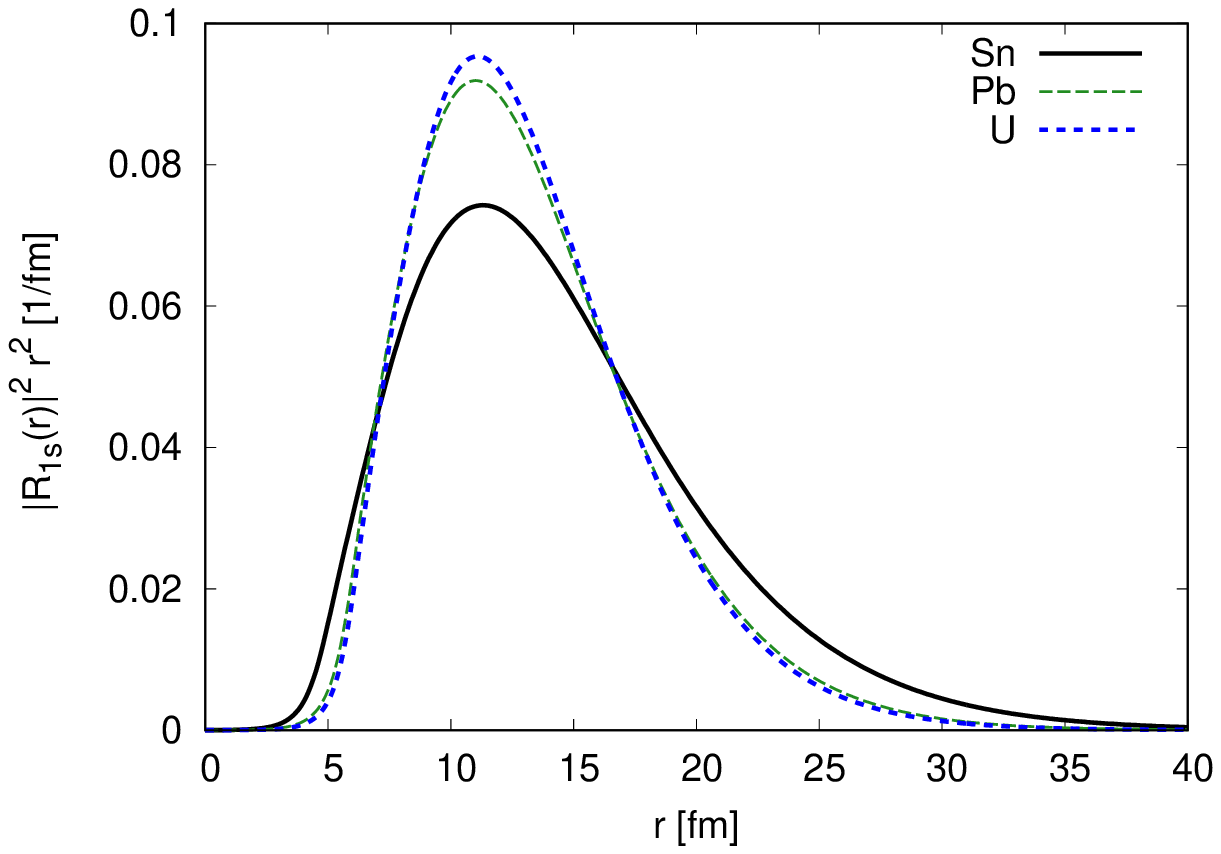}
\caption{ Radial density distribution $|R_{1s}(r)|^2 r^2$ of single pionic 1$s$ atoms of $^{121}$Sn, $^{207}$Pb and $^{237}$U plotted as functions of $r$.
\label{wf_Sn.Pb.U}}
\centering
\includegraphics[width=0.6\textwidth]{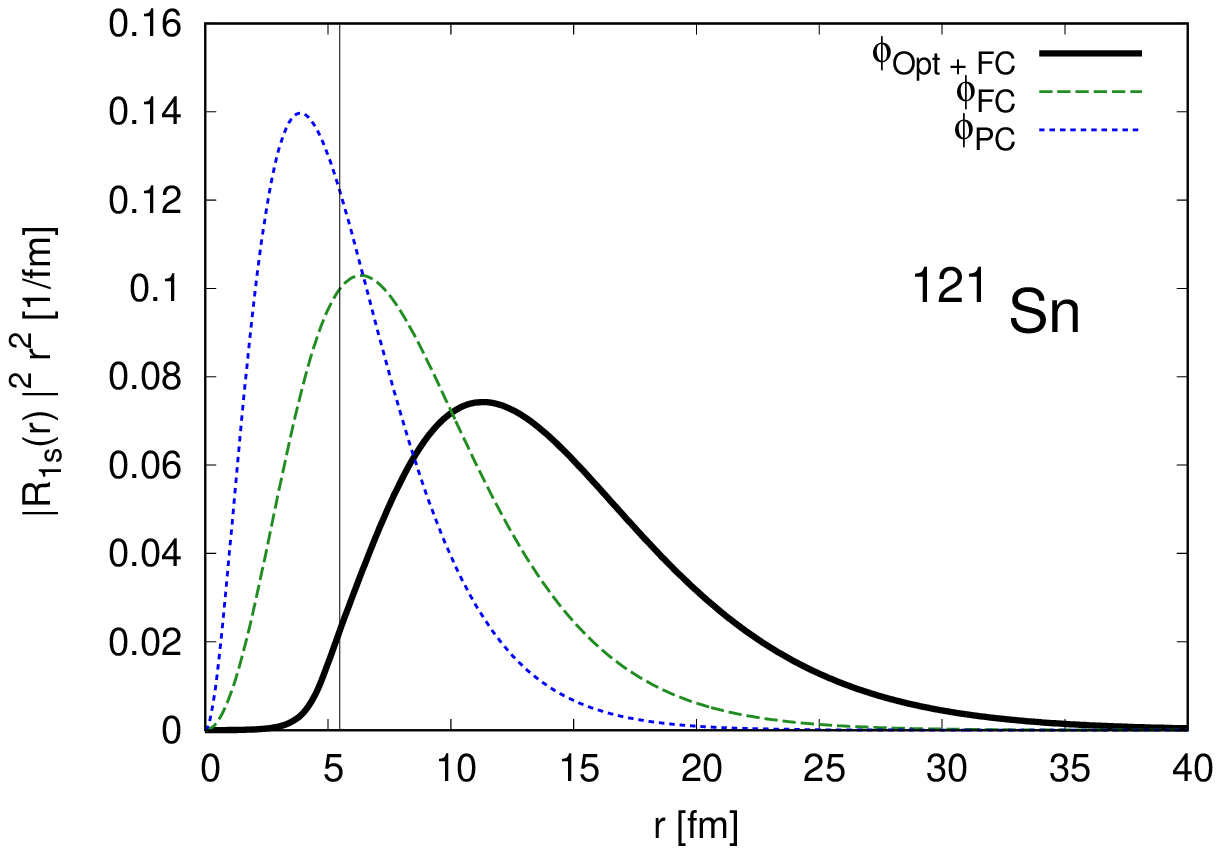}
\caption{Radial density distribution $|R_{1s}(r)|^2 r^2$ of single pionic 1$s$  atom in $^{121}$Sn plotted as functions of $r$.  The solid line (${\rm Opt+FC}$) indicates the pion density
obtained by solving the Klein-Gordon equation (Eq.~(\ref{KGeq}))
with the optical potential $V_{\rm Opt}$ and the finite-size Coulomb potential $V_{\rm FC}$, and the dashed line (${\rm FC}$) indicates the density only with the finite-size Coulomb potential $V_{\rm FC}$.
The distribution obtained by the Schr\"odinger equation with the point Coulomb potential $V_{\rm PC}$ is also shown by the dotted line (${\rm PC}$) for comparison. The vertical line indicates the nuclear radius of $^{121}$Sn.
\label{wf_121Sn}}
%\end{figure}
%\begin{figure}[htb]
\end{figure}

 Now, we will see the effects of the energy shift due to the $\pi\pi$ interactions by the first-order perturbative theory.
In Table~\ref{Eshift_table}, we show the calculated energy shifts of the ground state ($1s,1s$) of the double pionic atom due to the $\pi\pi$ electromagnetic interaction ($\Delta E_{\rm em}$) and the strong interaction ($\Delta E_{\rm s}$)
 evaluated by the perturbation theory.

We first look at the results for the realistic wave function obtained by solving Eq.~(\ref{KGeq}) with the optical potential $V_{\rm opt}$ and with the finite Coulomb potential $V_{\rm FC}$. By comparing the size of the shifts by the electromagnetic interaction and strong interaction, we find that
the value of the electromagnetic interaction shift $\Delta E_{\rm em}$ is larger than those of the strong interaction shift $\Delta E_{\rm s}$ 
because the $\pi \pi$ electromagnetic interaction $V^{\rm em}_{\pi \pi}$ is more effective in such a long
range than the $\pi \pi$  strong interaction $V^{\rm s}_{\pi \pi}$ that the averaged relative distance 
of two pions is calculated to be around 20~ fm for the $(1s,1s)$ state in Sn.
%
%
%-- Sn case;  delta to gauss no chigai  
In Table~\ref{Eshift_table}, we also show the calculated energy shifts $\Delta E_{\rm s}$
for the two different spatial distributions of the potential $V^{\rm s}_{\pi \pi}$, 
the $\delta$ function of Eq.~(\ref{V_delta}) and the Gaussian form of Eq.~(\ref{V_gauss}).
We find that the difference of the energy shifts by the two different forms of the potential is small.
Thus, in the following results we will only consider the strong energy shift $\Delta E_{\rm s}$ by the $\delta$ function potential.

We also show in Table~\ref{Eshift_table} 
the energy shifts $\Delta E_{\rm em}$ and $\Delta E_{\rm s}$ in the different nuclei $^{207}$Pb and $^{237}$U.
In the heavier nucleus, the single pionic binding energy $B_{\pi}$ becomes larger, 
such as  $B_{\pi}^{1s} = 6933.7$~keV in $^{207}$Pb
and $B_{\pi}^{1s}= 7848.0 $~keV in $^{237}$U,
because the attractive Coulomb potential $V_{\rm FC}$ becomes stronger.
As we can see in Fig.~\ref{wf_Sn.Pb.U}, where these pionic $1s$ wave functions are plotted, the radial distributions are more compact for the heavier nuclei. Thus, the energy shifts for heavier nuclei tend to be larger.

%--- density dependent
Here, in order to  %It should be noted here that we 
see the possible change of the $V_0$ value in the nucleus, 
we evaluated the energy shifts by using the density dependent $V_0$ as described in Eqs.~(\ref{f_pi_rho}) and (\ref{V0_rho}).
The effect of the density dependence is very small in all nuclear cases with the realistic wave function $\phi_{\rm Opt + FC}$. 
The effect is a little bit larger in the heavier nuclei cases such as Pb and U
as we have expected.
Thus, it is found to be very difficult to obtain the information on density dependence of the $\pi\pi$ interaction (scattering length of $I=2$ channel) from the double pionic atoms.

%--- Other wf
To understand these results in detail,
we additionally consider two other single pionic wave functions in $^{121}$Sn to evaluate the energy shifts. In Fig.~\ref{wf_121Sn}, we show the radial density distributions of single pionic 1$s$ states in $^{121}$Sn for three different potential.
The densities indicated as (Opt+FC) and (FC) are obtained by solving the Klein-Gordon equation (Eq.~(\ref{KGeq})) with the optical potential $V_{\rm opt}$ and the finite-size Coulomb potential $V_{\rm FC}$ ({\rm Opt+FC}),
and with the finite-size Coulomb potential $V_{\rm FC}$ only ({\rm FC}).
We also show the density distribution of the non-relativistic pionic atom of the point Coulomb potential $V_{\rm PC}$.
We can see in Table~\ref{Eshift_table} that the energy shifts with $\phi_{\rm PC}$ is the largest within all cases considered here.
This is because the wave function $\phi_{\rm PC}$ is well localized
inside of the nucleus and two pions have the largest overlap of density.
%%%%%%%%%%%
On the other hand, since the pion-nucleus optical potential $V_{\rm opt}$ is repulsive
especially for $s$-states, the wave function ($\phi_{\rm Opt + FC}$) is pushed  away
from the nucleus as shown in Fig.~\ref{wf_121Sn}.
Therefore, the energy shifts for $\phi_{\rm Opt + FC}$ are found to be smaller than those in the other two cases.

\begin{table*}[tb]
\centering
\caption{\label{Eshift_table2} 
Energy shifts shown for five different combinations of two pionic states ($1s,1s$), ($1s,2s$), ($1s,2s$), ($2p,2s$), ($2s, 2s$)  in $^{121}$Sn due to the $\pi\pi$ electromagnetic interaction ($\Delta E_{\rm em}$) and the strong interaction ($\Delta E_{\rm s}$) obtained by the first-order perturbation theory. } 
\begin{tabular}{ccccccccc}
\hline
\hline
     & $(1s,1s )$ &  $(1s, 2p)$ & $(1s,2s)$ & $(2p,2s)$ & $(2s,2s)$ \\         
\hline
 $\Delta E_{\rm em}$ [keV] &  91.0 & 89.1  & 47.2 & 43.8  & 33.8  \\
 $\Delta E_s$ [keV] & 7.5  & 8.0  & 2.2 & 1.3  & 0.48  \\
\hline
\hline
\end{tabular}
%\end{table*}
%-------------------------------------
%\begin{table*}[htb]
\centering
\caption{\label{Eshift_table3} 
Energy shifts $\Delta E_{\rm em}$ and $\Delta E_{\rm s}$ of the double pionic $(2p, 2p)$ state with the total angular momentum $L$.}
\begin{tabular}{cccccccc}
\hline
\hline
 $L$ &  0 & 1 & 2  \\         
\hline
 $\Delta E_{\rm em}$ [keV] &  77.8 & 58.5 & 66.2   \\
 $\Delta E_s$ [keV]        & 8.0   & 0.0  & 3.2  \\
\hline
\hline
\end{tabular}
\end{table*}

\begin{figure}[!htb]
\centering
\includegraphics[width=0.52\textwidth]{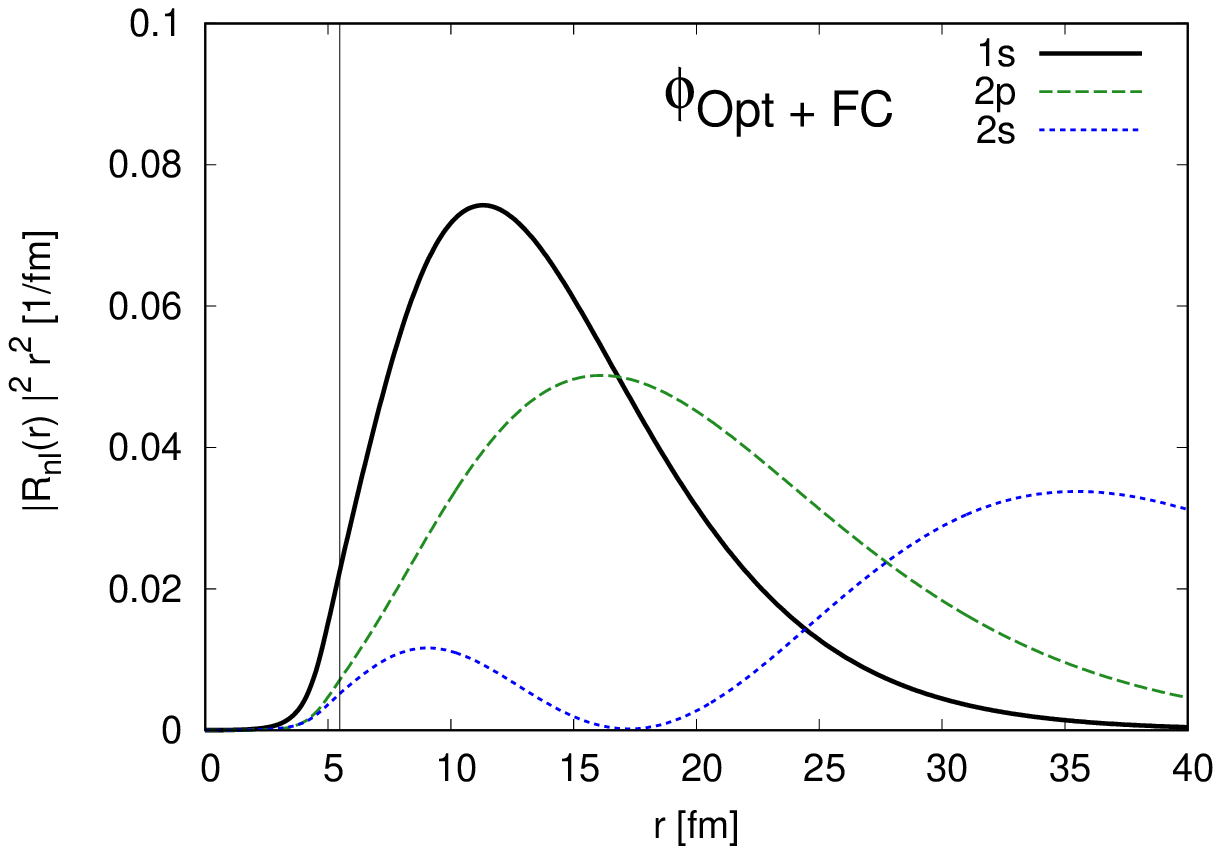}
\caption{
Radial density distribution $|R_{nl}(r)|^2 r^2$ of single pionic 1$s$, 2$p$ and 2$s$ atoms of $^{121}$Sn obtained with the optical and finite size Coulomb potential are plotted as functions of $r$. The vertical line indicates the nuclear radius.
\label{wf_121Sn_various}}
%\end{figure}
%\begin{figure}[tb]
\centering
\includegraphics[width=0.52\textwidth]{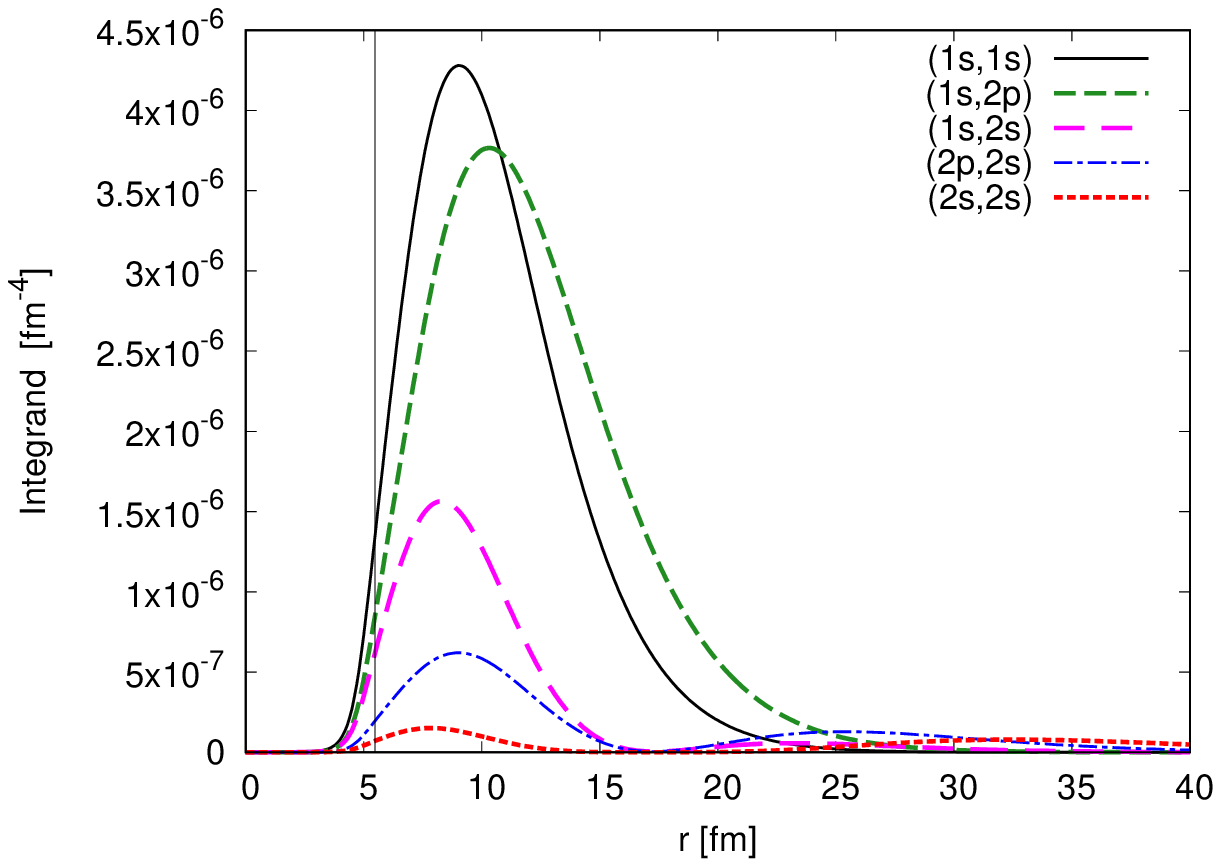}
\caption{The integrands appearing in the calculation of the strong energy shift $\Delta E_{s}$ for the five different combinations of two pionic states as a function of $r$. The potential strength $V_0$ is not included. The vertical line indicates the nuclear radius.}
\label{fig:Es_wf2}
\centering
\includegraphics[width=0.52\textwidth]{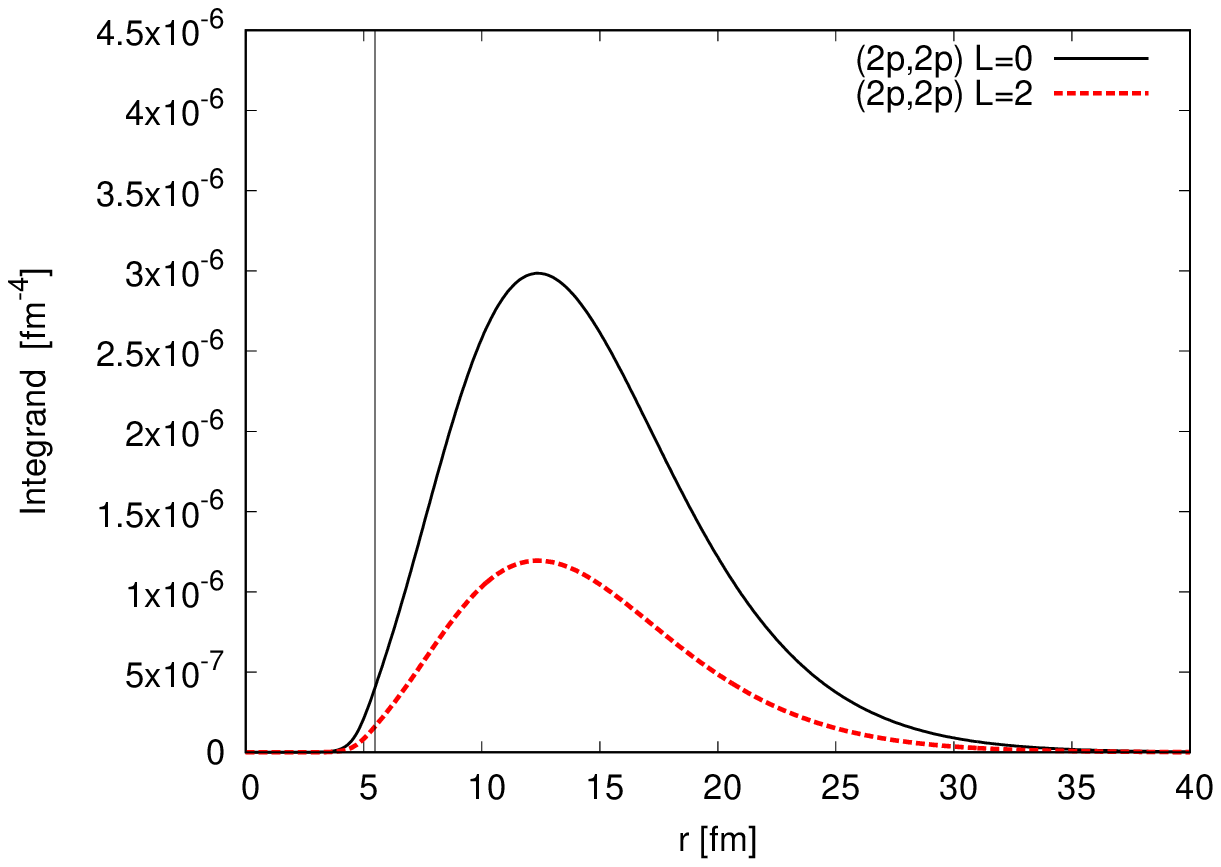}
\caption{Same as Fig.~\ref{fig:Es_wf2} for the double pionic $(2p, 2p)$ state with the total angular momentum $L=0$ and 2. 
\label{fig:Es_wf2_2p2p}}
\end{figure}

In Tables~\ref{Eshift_table2} and~\ref{Eshift_table3}, we summarize the calculated energy shifts $\Delta E_{\rm em}$ and $\Delta E_{s}$ for the different combinations of two pionic states in $^{121}$Sn.
We have used the realistic single pionic $1s$, $2p$ and $2s$ wave functions ($\phi_{\rm Opt + FC}$) as shown in Fig.~\ref{wf_121Sn_various}.
Similar to the results shown in Table~\ref{Eshift_table},
the energy shifts $\Delta E_{\rm em}$ are larger than those of $\Delta E_{\rm s}$ for all  combinations of two pionic states.
The size of the energy shifts shown in Tables~\ref{Eshift_table2} and~\ref{Eshift_table3} are expected to be determined by the overlap of the pionic wave functions as explained above. Thus, to get an intuitive idea of the shift size, we plot in Fig.~\ref{fig:Es_wf2} the integrands appearing in the calculations of the energy shifts $\Delta E_{s}$ in Table~\ref{Eshift_table2}, without the potential strength $V_0$. For example, for the $(1s,1s)$ case in Fig.~\ref{fig:Es_wf2}, the function of $\frac{1}{4\pi} r^2 |R_{10}(r)|^4$ is plotted corresponding to Eq.~(\ref{eq:Es_1s1s}) in Appendix~\ref{appA_s}. 
From the behavior of the integrands, we can understand the size of $\Delta E_{s}$ in Table~\ref{Eshift_table2} intuitively except for the $(1s,2p)$ combination. Because the 1$s$ wave function is more compact than the $2p$ one, we would expect $\Delta E_{\rm em}$ and $\Delta E_{s}$ to be smaller in the $(1s, 2p)$ case than in the $(1s,1s)$ case. Actually, $\Delta E_{\rm em}$ are practically the same and $\Delta E_{s}$ is even bigger in $(1s, 2p)$ than in $(1s,1s)$. The reason must be seen in the interference term in Eqs.~(\ref{Eshift_em}) and (\ref{eq:Eshift_s}) using Eq.~(\ref{WF_sp}). This is an effect of the Bose statistics and is the same effect that makes the parahelium having symmetric orbital wave functions less bound than the orthohelium.

For the double pionic $(2p, 2p)$ state, we have three different total angular momentum states $L=0, 1, 2$. The energy shifts for these states are shown in Table~\ref{Eshift_table3}. In both energy shifts $\Delta E_{\rm em}$ and $\Delta E_{s}$, the results for the $L=0$ state are the largest, while the results for the $L=1$ state are the smallest. The energy shift $\Delta E_{s}$ of the $L=0$ state is $8.0$~keV and is as large as that of $(1s,2p)$ state. 
This value is the largest $\Delta E_{s}$ among all cases considered for $^{121}$Sn with the realistic wave function $\phi_{\rm Opt + FC}$. 
In Fig.~\ref{fig:Es_wf2_2p2p} we plot again the integrands for the double pionic $(2p, 2p)$ states as Fig.~\ref{fig:Es_wf2} with $L=0$ and 2. 
For $L=0$, for example, the function $\frac{3}{4\pi} r^2 |R_{21}(r)|^4$ is plotted in accordance with Eq.~(\ref{Es_2p2p_L0}) in Appendix~\ref{appA_s}.
From Figs.~\ref{fig:Es_wf2} and \ref{fig:Es_wf2_2p2p} we see that the integrand for the $(2p, 2p)$ state has relatively wider distribution than those of the $(1s,1s)$ and $(1s,2p)$ states as expected from the wave functions in Fig.~\ref{wf_121Sn_various}.
The $\Delta E_{s} =0$ for $L=1$ is a consequence of the antisymmetry of the wave function which vanishes at $\vec{r}_1 = \vec{r}_2$ and, hence, the $\delta(\vec{r_1}-\vec{r_2})$ function of Eq.~(\ref{V_delta}) gives a null contribution.

\section{Summary}\label{sec:summary}
We have studied theoretically the structure of the double pionic atoms for the six different combinations of two pionic states, $(1s,1s)$, $(1s,2p)$, $(1s,2s)$ $(2p,2s)$, $(2s,2s)$, and $(2p,2p)$ in $^{121}$Sn.
The deeply bound single pionic $1s$, $2s$, $2p$ states have been experimentally observed in heavy nuclei. 
We also studied the double pionic atoms in the cases of the heavier nuclei and the cases with the density dependent $\pi \pi$ interaction.
We have evaluated the energy shifts by the $\pi\pi$ interactions using the perturbation theory with the realistic wave functions of the single pionic atoms.
For the strong interaction between two pions, we assume the delta function and the Gaussian form as the spatial distribution of the potential. The potential strength and range are fixed to reproduce the $\pi\pi$ scattering length.

We found that the energy shifts due to the $\pi\pi$ electromagnetic interaction are  33.8--91.0~keV for the states considered here for the Sn nucleus. 
We also found that the energy shifts due to the $\pi\pi$ strong interaction is 0--8.0~keV for the Sn states.
The energy shifts of the double pionic ($1s,2p$) and ($2p, 2p$) states are relatively large and of similar size to that of the ground ($1s,1s$) state.
The energy shifts tend to be larger for the heavier nuclei because of more compact pion distribution and larger overlap.
Actually by comparing the results with Coulomb wave functions ($\phi_{\rm FC}$ and $\phi_{\rm PC}$), we found that the size of the energy shifts are much reduced by the realistic $\pi$-nucleus potential effects which push outwards the pion wave functions and suppress the overlap between them.
Obtained energy shifts of the double pionic states due to the $\pi\pi$ interaction are the first quantitative results and will be the clue for the further studies of the structure of the multi pionic atoms.   

Recently, the studies of the single deeply bound pionic atoms are developed theoretically and experimentally.
In the recent experiment of the deeply bound single pionic atom observation, the errors of the experimentally measured binding-energies are even smaller than 10~keV~\cite{Nishi2}.  
While the ($d,^3$He) reaction is used for producing single pionic atoms, secondary pion beams should be employed to populate two pions~\cite{formation2pion,Nieves:1992kc}. Kinematically, recoilless condition is achieved for formation of double pionic atoms by ($\pi^-,p$) reaction with the incident momentum of about 300 MeV/$c$~\cite{Nieves:1992kc}, which is available at several facilities in the world, for instance, PSI and J-PARC. In order to discuss experimental feasibility quantitatively, a elaborate calculation of the formation spectrum, as well as an evaluation of the background cross sections, is mandatory. We would leave it for future work.

% \section{Conclusion}
% The conclusion text goes here.

\section*{Acknowledgment}
N. I. thanks Professor E. Oset for useful discussions and valuable comments.
 This work was supported by JSPS Overseas Research Fellowships and JSPS KAKENHI Grants No. JP19K14709, No.~JP16K05355 and No.~JP17K05449. %,  and No.~****.

%Insert the Acknowledgment text here.


%\bibliographystyle{ptephy}
%\bibliography{sample}
%
% once the .bbl file has been generated then place the text in your article.

\appendix
\section{Expressions of the energy shifts for double pionic states}\label{appA}
We summarize the formulation to calculate the energy shifts due to the $\pi\pi$ interaction by perturbation theory.

\subsection{Energy shift $\Delta E_{\rm s}$ by the $\pi\pi$ strong interaction $V^{\rm s}_{\pi \pi}$ }\label{appA_s}
To obtain the energy shift $\Delta E_{\rm s}$ by the $\pi\pi$ strong interaction $V^{\rm s}_{\pi \pi}$, we calculate Eq.~(\ref{eq:Eshift_s}). 
First, we show the formalism for the interaction with a $\delta$ function as in Eq.~(\ref{V_delta}) for the $\pi\pi$ strong interaction $V^{\rm s}_{\pi \pi}$.
For the ground state $(1s,1s)$ of the double pionic atom, the energy shift can be written using the wave function of Eq.~(\ref{WF_ground}) as, 
\begin{eqnarray}
\Delta E_{\rm s}
&=& \langle \Phi_0 (\vec{r}_1, \vec{r}_2) 
| V_0 \delta ({\vec{r}_1}- {\vec{r}_2}) |
\Phi_0 (\vec{r}_1, \vec{r}_2) \rangle
\nonumber\\
&=& 
\int d\vec{r}_1 d\vec{r}_2 
\phi^{*}_{100}(\vec{r}_1) \phi^{*}_{100}(\vec{r}_2) 
V_0 \delta({\vec{r}_1}- {\vec{r}_2}) 
\phi_{100}(\vec{r}_1) \phi_{100}(\vec{r}_2)  
\nonumber\\
&=&
\frac{1}{4\pi}
\int r_1^2 dr_1  V_0 
\left|R_{1 0}(r_1) \right|^4,
\label{eq:Es_1s1s}
\end{eqnarray}
where we have used $Y_{0}^{0}(\hat r) = \frac{1}{\sqrt{4\pi}}$.
The energy shift $\Delta E_{\rm s}$ for the $(2s,2s)$ state can be evaluated by the exactly same expression as Eq.~(\ref{eq:Es_1s1s}) except for the radial wave function $R_{2 0}(r)$. 

The energy shift $\Delta E_{\rm s}$ for the $(1s,2p)$ state can be evaluated by using the wave function of Eq.~(\ref{WF_sp}) as, 
\begin{eqnarray}
\Delta E_{\rm s}
&=& 
\int d\vec{r}_1 d\vec{r}_2 
\frac{1}{\sqrt{2}}[
\phi_{100}^{*}(\vec{r}_1) \phi_{21m}^{*}(\vec{r}_2)  
+
\phi_{21m}^{*}(\vec{r}_1) \phi_{100}^{*}(\vec{r}_2) 
]
\ V_0 \delta({\vec{r}_1}- {\vec{r}_2})  \nonumber\\
& & \times
\frac{1}{\sqrt{2}}[
\phi_{100}(\vec{r}_1) \phi_{21m}(\vec{r}_2)  
+
\phi_{21m}(\vec{r}_1) \phi_{100}(\vec{r}_2) 
] \nonumber\\
&=&
\frac{1}{2\pi}  \int r_1^2 dr_1   V_0 
\left|R_{10}(r_1) R_{21}(r_1) \right|^2,
\label{eq:Es_1s2p}
\end{eqnarray}
where we have used $\int | Y_{0}^{0}(\hat r) Y_{1}^m (\hat r)|^2 d \Omega = \frac{1}{4\pi}$.
The energy shifts $\Delta E_{\rm s}$ for the $(1s,2s)$ states can be evaluate as Eq.~(\ref{eq:Es_1s2p}) by using $R_{20}$ instead of $R_{21}$. The shift $\Delta E_{\rm s}$ for the $(2s,2p)$ state is also calculated by the Eq.~(\ref{eq:Es_1s2p}) by using $R_{20}$ instead of $R_{10}$.

%*($2p, 2p$):
For the double pionic $(2p,2p)$ states, the total angular momentum $L$ can be $L=0$, 1, and 2.   
We need to evaluate the energy shifts for three states with different $L$.
Using Eqs.~(\ref{eq:WF_2p2p}) and (\ref{eq:WF_2p2p_2}), the wave function of the state with $L=M=0$ is written as 
\begin{equation}
\Phi_{0 0}(\vec{r}_1, \vec{r}_2)  
=  \sqrt{\frac{1}{3}}  \left[ \phi_{21 -1}(\vec{r}_1) \phi_{211}(\vec{r}_2) 
-   \phi_{21 0}(\vec{r}_1) \phi_{210}(\vec{r}_2) 
+   \phi_{211}(\vec{r}_1) \phi_{21-1}(\vec{r}_2)  \right],
\end{equation}
then the energy shift is evaluated as
\begin{eqnarray}
\Delta E_{\rm s}
= \frac{3}{4\pi}  \int r_1^2 dr_1   V_0 
\left|R_{21}(r_1) \right|^4.
\label{Es_2p2p_L0}
\end{eqnarray}
For the states with $L=1,2$, the energy shifts do not depend on the quantum number $M$.
For $L=1$ state, the wave function with $M=-1$, for example, can be written as,
\begin{equation}
\Phi_{1 -1}(\vec{r}_1, \vec{r}_2)  
=  -\sqrt{\frac{1}{2}}  \left[ \phi_{21 -1}(\vec{r}_1) \phi_{210}(\vec{r}_2) 
-   \phi_{210}(\vec{r}_1) \phi_{21-1}(\vec{r}_2)  \right].
\end{equation}
The energy shift $ \Delta E_{\rm s}$ for this wave function is 0. For other states with $M=0$ and 1, the values of $ \Delta E_{\rm s}$ are equal to 0 for $L=1$.
In the case of $L=2$, the wave function of the state with $M=-2$, for example, is written as, 
\begin{equation}
\Phi_{2 -2}(\vec{r}_1, \vec{r}_2) = 
\phi_{21 -1}(\vec{r}_1) \phi_{21-1}(\vec{r}_2). 
\end{equation}
The energy shift is calculated by the same expression of Eq.~(\ref{Es_2p2p_L0}) except for the factor $\frac{3}{10\pi}$ instead of $\frac{3}{4\pi}$. The results are same for all other $M$ states for $L=2$.

Finally, we also show the formula for the energy shifts $\Delta E_{\rm s}$ by the $\pi\pi$ strong interaction $V^{\rm s}_{\pi \pi}$ of the Gaussian form defined in Eq.~(\ref{V_gauss}), for the ground ($1s,1s$) state.
The energy shift $\Delta E_{\rm s}$ can be evaluated as, 
\begin{eqnarray}
\Delta E_{\rm s}
&=& 
\int d\vec{r}_1 d\vec{r}_2 
|\phi_{100}(\vec{r}_1) \phi_{100}(\vec{r}_2) |^2
 \frac{V_0 }{(b^2 \pi)^{3/2}} 
\exp \left[- \frac{ (\vec{r}_1 - \vec{r}_2)^2}{b^2} \right] 
\nonumber\\
& =&\int r_1^2 d r_1  r_2^2 d r_2 
|R_{10}(r_1) R_{10}(r_2) |^2
 \frac{V_0 }{(b^2 \pi)^{3/2}} 
\exp \left[- \frac{ r_1^2 + r_2^2}{b^2} \right] 
\sqrt{\frac{\pi}{ \frac{4r_1 r_2}{b^2}}} I_{1/2}\left(\frac{2r_1 r_2}{b^2}\right),
\nonumber\\
\label{Es_gauss_1s1s}
\end{eqnarray}
where $e^{\frac{  2\vec{r}_1 \cdot \vec{r}_2 } {b^2} }$ is expanded as,
\begin{eqnarray}
e^{\frac{  2\vec{r}_1 \cdot \vec{r}_2 } {b^2} } 
&=&
4\pi  \sum_{\ell=0}^{\infty} \sum_{m= -\ell}^{+ \ell} 
i_{\ell} \left(\frac{2r_1 r_2}{b^2}\right)
Y_{\ell}^{m*}(\hat{r}_1) Y_{\ell}^{m}(\hat{r}_2)
 \nonumber\\
&=&
4\pi  \sum_{\ell=0}^{\infty} \sum_{m= -\ell}^{+ \ell} 
\sqrt{\frac{\pi}{2 \frac{2r_1 r_2}{b^2}}} I_{\ell+1/2}\left(\frac{2r_1 r_2}{b^2}\right)
Y_{\ell}^{m*}(\hat{r}_1) Y_{\ell}^{m}(\hat{r}_2),
\end{eqnarray}
with the modified spherical Bessel function $i_{\ell}(x)$ which is related the Bessel function $i_\ell (x) =(-i)^\ell j_\ell (ix)$. In the $\ell =0$ case, it is written as,
\begin{equation}
 i_{0}(x) = \sqrt{\frac{\pi}{2x}} I_{1/2}(x) =\frac{\sinh x}{x}.
\end{equation}

\subsection{Energy shift $\Delta E_{\rm em}$ by the $\pi\pi$ electromagnetic interaction $V^{\rm em}_{\pi \pi}$}
To obtain the energy shift $\Delta E_{\rm em}$ by the $\pi\pi$ electromagnetic interaction $V^{\rm em}_{\pi \pi}$, we evaluate the expectation value defined in Eq.~(\ref{Eshift_em}). 
For the ground state $(1s,1s)$ of the double pionic atom, the energy shift can be evaluated by using the wave function of Eq.~(\ref{WF_ground}) as, 
\begin{eqnarray}
\Delta E_{\rm em}
=
\frac{e^2}{4 \pi \epsilon_0} \int_{0}^{\infty} r_1 dr_1 
|R_{1 0}(r_1)|^2
\left\{ \int_{0}^{r_1} r_2^2 dr_2  |R_{1 0}(r_2)|^2
+ r_1 \int_{r_1}^{\infty} r_2 dr_2 |R_{1 0}(r_2)|^2 \right\}.
 \nonumber\\
\label{eq:Eem_1s1s}
\end{eqnarray}
The energy shift $\Delta E_{\rm em}$ for the $(2s,2s)$ state can be evaluated by the same expression as Eq.~(\ref{eq:Eem_1s1s}) except for the wave function $R_{2 0}$ instead of $R_{10}$.

As for the ($1s,2s$) state, by using the Eqs.~(\ref{Eshift_em}) and (\ref{WF_1s2s}),
the energy shift $\Delta E_{\rm em}$ can be evaluated as,
\begin{eqnarray}
\Delta E_{\rm em}
&=&
\frac{e^2}{4 \pi \epsilon_0}  \int_{0}^{\infty} r_1 dr_1 
|R_{1 0}(r_1)|^2
\left\{ \int_{0}^{r_1} r_2^2 dr_2  |R_{2 0}(r_2)|^2
+ r_1 \int_{r_1}^{\infty} r_2 dr_2 |R_{2 0}(r_2)|^2 \right\}
\nonumber\\
&+& 
\frac{e^2}{4 \pi \epsilon_0} \int_{0}^{\infty} r_1 dr_1 
R_{10}^{*}(r_1) R_{2 0}(r_1)
\left\{ \int_{0}^{r_1} r_2^2 dr_2  R_{2 0}^{*}(r_2) R_{10}(r_2)
+ r_1 \int_{r_1}^{\infty} r_2 dr_2 R_{2 0}^{*}(r_2) R_{10}(r_2) \right\}.
\nonumber\\
\label{Eem_1s2s}
\end{eqnarray}

The energy shift for ($1s,2p$) state is evaluated as,
\begin{eqnarray}
\Delta E_{\rm em}
&=&  \frac{e^2}{4 \pi \epsilon_0} \int_{0}^{\infty} r_1  dr_1 |R_{10}(r_1)|^2
\Big\{ \int_{0}^{r_1} r_2^2 dr_2 |R_{21}(r_2)|^2
+
r_1 \int_{r_1}^{\infty} r_2 dr_2 |R_{21}(r_2)|^2
  \Big\}
\nonumber\\ 
&+ & 
\frac{e^2}{4 \pi \epsilon_0} \frac{1}{3}  \int_{0}^{\infty}  dr_1 R^*_{10}(r_1) R_{21}(r_1)
\Big\{ \int_{0}^{r_1}  r^3_{2}  dr_2  R^*_{21}(r_2) R_{10}(r_2)
+
r_{1}^3 \int_{r_1}^{\infty} dr_2 R^*_{21}(r_2) R_{10}(r_2)  
 \Big\}.
\nonumber\\ 
\label{eq:Eem_1s2p}
\end{eqnarray}
The factor of $\frac{1}{3}$ comes from the angular integration.
The energy shift $\Delta E_{\rm em}$ for the $(2p,2s)$ state can be evaluated by the same expression Eq.~(\ref{eq:Eem_1s2p}) except for the wave function $R_{20}$ instead of $R_{10}$.

% *($2p, 2p$) ***********
Finally, we show the formula of the energy shifts for the ($2p, 2p$) states, which are independent from the quantum numbers $M$ for the total angular momentum $L$.
For the $L=0$ state, the energy shift is evaluated as,
\begin{eqnarray}
\Delta E_{\rm em}
&=&
\frac{e^2}{4 \pi \epsilon_0} \int_{0}^{\infty} r_1 dr_1 
|R_{21}(r_1)|^2
\nonumber\\
&\times &
\left[ \int_{0}^{r_1} r_2^2 dr_2  |R_{21}(r_2)|^2 
\left\{ 1 + \frac{2}{5} \left(\frac{r_2}{r_1} \right)^2 \right\}
+ r_1 \int_{r_1}^{\infty} r_2 dr_2 |R_{21}(r_2)|^2 
\left\{ 1 + \frac{2}{5} \left(\frac{r_1}{r_2} \right)^2 \right\}
\right].
 \nonumber\\
\label{eq:Eem_2p2p}
\end{eqnarray}
For the $L=1$ states, we can evaluate the $\Delta E_{\rm em}$ using Eq.~(\ref{eq:Eem_2p2p}) changing the factors $\frac{2}{5}$ to $- \frac{1}{5}$.
Similarly, for $L=2$ states, we change the factors $\frac{2}{5}$ to $\frac{1}{25}$ in Eq.~(\ref{eq:Eem_2p2p}).

\end{document}